\documentclass[onecolumn]{aa} 
\usepackage{graphicx}
\usepackage{lscape}
\usepackage{txfonts}
\begin{document}
   \title{The star cluster system of the luminous elliptical galaxy NGC 1600}


   \author{B. X. Santiago
          \inst{1}
          }

   \offprints{B. Santiago}

   \institute{Instituto de F\'\i sica, Universidade Federal do Rio Grande\\ 
              do Sul, Av. Bento Gon\c calves, 9500, CP 15051, Porto Alegre\\
              \email{santiago@if.ufrgs.br}}

   \date{Received April 4th, 2008; accepted August 22nd, 2008}

 
  \abstract
   {Luminous elliptical galaxies generally display a rich star cluster
   system, whose properties provide strong constraints on the physics of galaxy
   formation and evolution. Star cluster system studies, however, 
   concentrate on galaxies located in nearby or rich galaxy clusters.}
   {We acquired deep B and I images of NGC 1600, a luminous elliptical in a galaxy 
   group to study its star cluster system. The images were obtained with the 
   Optical Imager at the Southern Telescope
   for Astrophysical Research for an exposure time of 1.66hr in each filter.}
   {The sample selection incompleteness was assessed as
    a function of magnitude and image background level. Source counts
    were measured for different elliptical annuli from the centre of NGC
    1600, background subtracted, and fitted with a Gaussian function.
    Colour distributions were derived as a function of galactocentric
    distance for sources measured successfully in both filters.
    Typical ages and metallicities were estimated based on single
    stellar population models.}
   {A clear excess of point sources around NGC 1600 was found in relation to the 
   nearby field. The source counts were consistent with a 
   Gaussian distribution typical of other luminous ellipticals. The 
   luminosity function
   fits provided an estimate of the density of clusters at the different 
   annuli that could be integrated in solid angle, resulting in an
   estimated total population of $N_{GC} \simeq 2850$ star clusters. 
   This yielded a specific frequency of $S_N \simeq 1.6$. 
   The colour distributions show a hint of bimodality, especially
   at $\simeq 20$ kpc from the centre. Clusters in this
   region may be associated with a ring or shell perturbation.
   Finally, the star cluster candidates were cross-correlated to discrete X-ray
   sources and a coincidence rate of $\simeq 40\%$ was found. These are
   likely to be globular clusters harboring low-mass X-ray binaries.}


   \authorrunning{Santiago}

   \titlerunning{The star cluster system of NGC 1600}

   \maketitle
%

\section{Introduction}

Star cluster systems reveal significant information about their host galaxy
formation and evolution. By far, luminous ellipticals display the most 
conspicuous cluster systems and have been the popular targets for 
studies of star cluster populations (Kundu et al 1999, 
Harris et al 2006). Studies are also 
concentrated on nearby galaxies, some in the field, but most
often in galaxy clusters, such as Virgo and Fornax 
(Larsen et al 2001, Santiago \& Elson 1996, Kissler-Patig et al 1997, 
Peng et al 2006a, Strader et al 2006). An important result of 
detailed analyses of extragalactic clusters is that clusters in 
luminous elliptical galaxies present bimodal colour distributions. 
These are often interpreted
as evidence for separate episodes of star and cluster formation, often
involving galaxy merging (Ashman \& Zepf 1992, Cot\^e et al 1998;
see recent review by Brodie \& Strader 2006). Even though most
studies refer to extragalactic star cluster systems as {\it globular
cluster systems}, as in fact globular clusters dominate the cluster 
counts, the discovery of new classes of clusters, such
as {\it faint fuzzies} (Larsen \& Brodie 2000, Brodie \& Larsen 2002) 
and {\it diffuse star clusters} (Peng et al 2006b) suggests
that a broader terminology should be adopted.

Galaxies in lower density environments, such as galaxy groups,
have also been studied for the presence of extragalactic star clusters 
(i.e., Kundu \& Whitmore 2001, Chies-Santos et al 2006). However,  
many bright elliptical 
galaxies in groups still lack an investigation of their cluster system.
One such case is the luminous elliptical galaxy NGC 1600.

NGC 1600 is the most luminous elliptical in its loose 
galaxy group,
which also contains two other early-types: NGC 1601 and NGC 1603.
Optical surface photometry, both photographic 
(Barbon et al 1984) and CCD (Mahabal et al 1995) imaging were obtained
for the galaxy. Near infra-red surface photometry
was completed by Rembold et al (2002). These studies
revealed a flattened and boxy shape. Sivakoff et al (2004) obtained
Chandra X-ray images and identified dozens of point sources
superimposed on the diffuse emission. They tentatively associated the unresolved
sources, some of which are classified as ultra-luminous
in X-rays, with the population of globular clusters (GCs).

   \begin{figure}
   \centering
   \includegraphics[width=\textwidth]{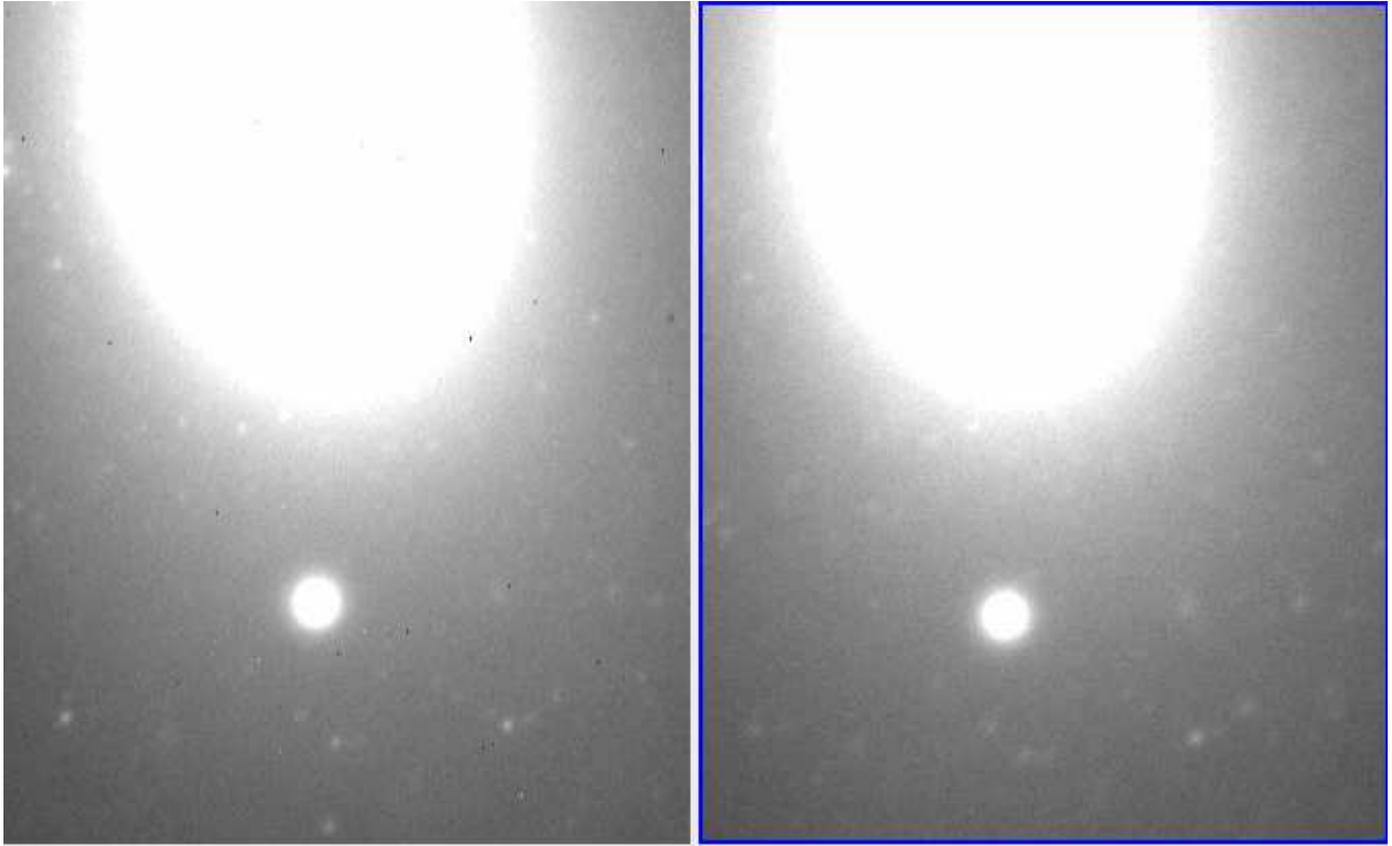}
   \caption{Close up on the southern side of NGC 1600 using the final 
            SOI/SOAR images. North is upwards and East is to the left of the
            images. Left panel: I band; right panel: B band.}
    \label{combBI}
    \end{figure}

In this paper, we attempt to detect the star cluster system of NGC 1600
using deep imaging with the Southern Telescope for Astrophysical 
Research (SOAR). Our primary goal is to estimate the total number
and specific frequency of clusters in NGC 1600, to derive their 
colour distribution and associate the star cluster sample with the X-ray
point sources identified by Sivakoff et al (2004). 


\section{Data and Photometry}

   Deep images of NGC 1600 were taken with the SOAR telescope in B and I
   filters. The SOAR Optical Imager (SOI) consists of two E2V CCDs,
   each one with 2k x 4k pixels, covering a field of view of 5.5x5.5 
   arcminutes. A 2x2 binning was used, yielding a detector scale of 0.154
   arcsec/pixel. A total of 6 x 1000s exposures (1.67 hr integration 
   time) were taken in slow read-out mode with each filter during two 
   nights. Photometric standards were also 
   observed on the same nights, along with bias and flat-field images.

   The individual exposures were bias-subtracted and flat-fielded using
   standard tasks from the Image Reduction and Analysis Facility
   (IRAF). The SOI I band images are affected by interference 
   that causes a large-scale fringing pattern. These were
   corrected using the IRAF mscred.rmfringe task, which
   computes the pattern amplitude minimizing the difference between 
   the target image and a {\it fringe pattern image} (usually blank
   sky), and corrects the former.
   The original Multiple Extension Format (MEF) files
   were converted into Flexible Image Transport System (FITS) files, 
   again using the IRAF mscred package. The 
   exposures from the same filter were aligned and combined into a 
   final image, which was used for sample selection and photometry. 
   Figure \ref{combBI} shows a close-up of the southern side of NGC 1600 
   of the final I and B combined SOAR/SOI images. 
   A large number of point sources are seen superimposed onto the galaxy's
   diffuse light. These are the star cluster candidates. The northern side of 
   the image is partly affected by a strong diffraction spike from a very 
   bright star. Also present in the SOI field (but not in 
   Fig. \ref{combBI}) are the two other galaxies belonging to the same 
   group, NGC 1603 and NGC 1601. 

   The IRAF daophot package was used to identify point sources 
   and measure their aperture magnitudes. This was completed separately in
   each of the B and I combined images. Sources with a detection confidence
   interval 2.5$\sigma$ above the sky 
   background were automatically detected with daofind. An aperture of 
   4 pixels = 0.61 arcsec was adopted to measure the magnitudes in both 
   filters using the task phot from the same package. The magnitudes
   were measured over the combined image used for source selection but 
   after the diffuse light from NGC 1600 was modelled and
   subtracted-off. Isophote fitting with the IRAF task ellipse was
   applied to model the
   galaxy intensity distribution. This information was then used to 
   generate a model of NGC 1600 with the task bmodel. This model was 
   subtracted from the combined image to produce magnitude 
   measurements that were unaffected by the steep gradient caused by NGC 1600 light 
   profile. Since a low threshold above the background was adopted in 
   daofind, a visual inspection of the model subtracted image 
   revealed an insignificant amount of extra cluster candidates 
   relative to the authomatic detection. This is important, because  
   it is far easier to compute photometric incompleteness in a sample
   selected by an automatic procedure rather than by eye.
   
   Photometric calibration was based on 36 magnitude measurements in
   each filter for 7 standard stars taken from Landolt (1992) and 
   observed at different airmasses. The magnitudes were measured in the
   same aperture as the point sources in NGC 1600. The calibration fields used for this
   purpose were SA 95 and PG0231+051 listed in Landolt (1992). The 
   calibration transformation from 
   instrumental magnitudes into Johnson-Cousins system included a linear 
   $(B-I)$ colour term, a linear airmass term, and a zero-point. A least 
   squares fitting of the three coefficients was carried 
   out for each filter separately. Saturated stars and deviant 
   measurements were eliminated by applying a $2\sigma-$clipping 
   algorithm iteratively to the magnitude residuals, 
   until convergence. In the B band, a total of 11 measurements were 
   eliminated in the clipping process; for the remaining 25 measurements, 
   the final fit dispersion was $\sigma_B = 0.04$. In the I-band, 
   16 measurements were used in the final fit, with a dispersion 
   $\sigma_I =  0.06$.

   \begin{figure}
   \centering
   \includegraphics[width=\textwidth]{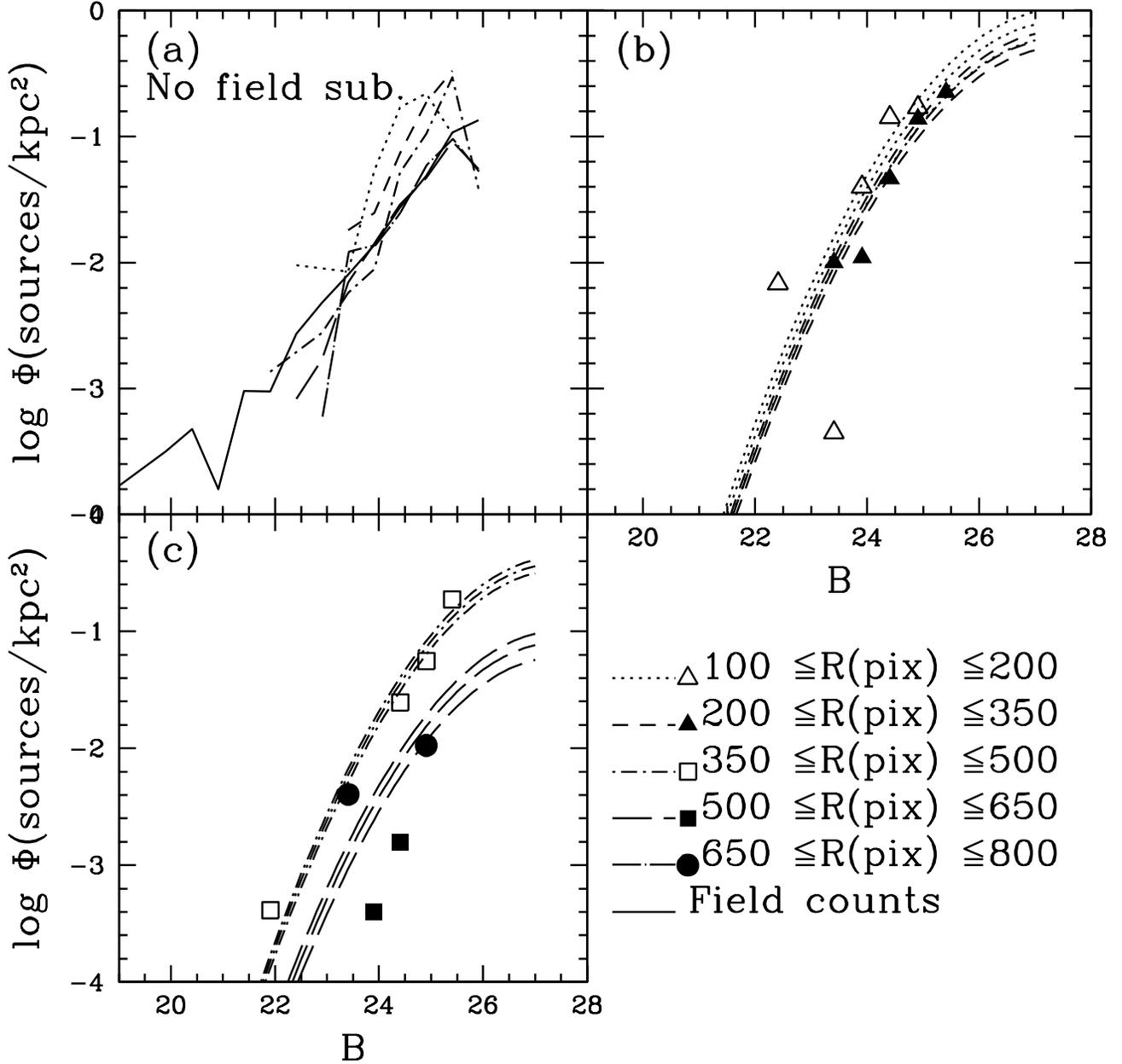}
   \caption{Panel {\it a)}: Total source counts as a function of 
            calibrated B magnitudes in different elliptical annuli 
            centred on NGC 1600. The solid line shows the
            source counts in the field region, away from NGC 1600 and 
            from the other galaxies in its group. The semi-major axis
            range (in pixels) of the different annuli is given in the 
            lower-right. 
            Panels {\it b,c)}: The symbols are the field-subtracted 
            counts in the same annuli as in panel {\it a}.
            The middle lines are Gaussian fits to the counts, using
            Eqs. (1) and (2). The lines immediately above and below of 
            each fit are Gaussian functions with a 1 $\sigma$
            variation in normalization relative to the fitted one. 
            The final fit, shown by a long dashed line
            in panel {\it c}, was carried out for the range $500 \leq 
            R(pix) \leq 800$}. 
    \label{lfs1600b}
    \end{figure}

   After calibration, magnitudes were corrected for foreground extinction
   in the direction towards NGC 1600. Schlegel et al (1998) quoted 
   $A_B=0.191$ and $A_I=0.09$. Burstein \& Heilles (1982) quoted $A_B=0.08$. The 
   average of both $A_B$ determinations was taken, yielding 
   $A_B=0.14$. 
   Correspondingly, $A_I = 0.065$. These values were used to 
   correct the calibrated magnitudes and colours, when available.

\subsection{Completeness analysis}

   Completeness experiments were performed on the combined images using
   the daophot.addstar task. Artificial stars spanning a wide range
   of B and I magnitudes were added to the combined images. These were 
   created by fitting a moffat point spread function (with $\beta
   =1.5$) fitted to bright, isolated, and non-saturated point sources 
   found in the images themselves. The fraction of recovered artificial 
   stars depended not only on their magnitudes but also on their
   location, and was lower in regions close to the centre of NGC 1600, 
   where the background was dominated by the galaxy light. The 
   effect of NGC 1603 and NGC 1601 on sampling 
   completeness in the B and I image was also modelled. The background 
   level was locally contaminated by a strong 
   diffraction spike towards the north of NGC 1600. This region
   was masked during the analysis, in addition to the saturated regions 
   in NGC 1600, NGC 1603 (in I only), and NGC 1601 (in both B and I 
   images). 

   From the completeness experiments, the completeness
   function could be modelled as a bivariate function of magnitude and 
   background level. This was achieved as follows: for each point source
   found in a given filter, we first determined, by cubic spline 
   fits to the completeness vs. magnitude relation, the completeness
   fraction at its measured magnitude for different 
   sky levels. This yielded a completeness versus sky level relation at the
   source magnitude. Another cubic spline was then applied to this 
   relation to evaluate completeness at the background 
   level at the location of the source. Each star was then assigned a 
   weight equal to the product $1/c(B)~\times~1/c(I)$ where $c(B)$ and $c(I)$ 
   are the completeness fractions calculated for the B and I images, 
   respectively.

   \begin{figure}
   \centering
   \includegraphics[width=\textwidth]{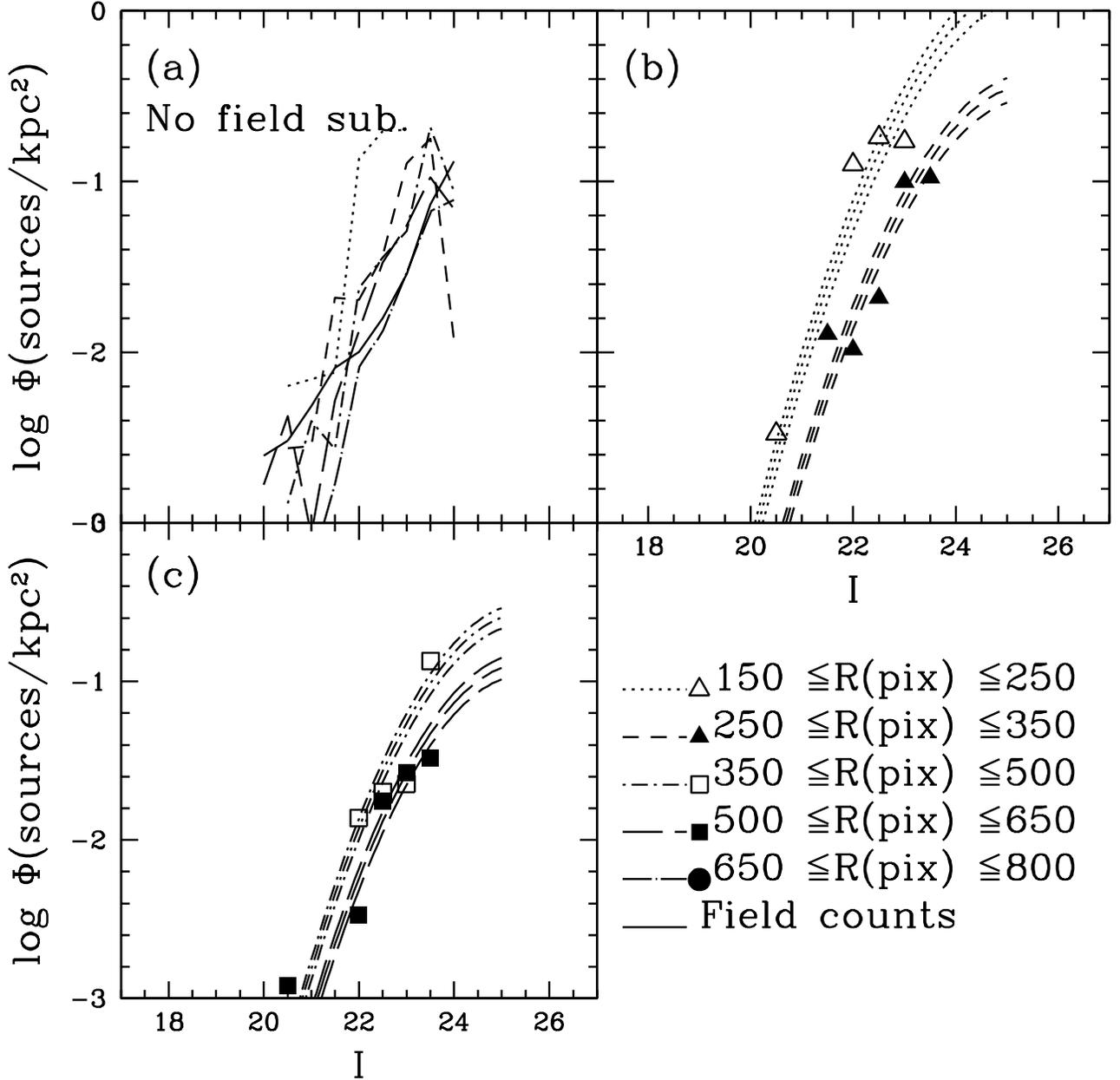}
   \caption{Panel {\it a)}: Total source counts as a function of 
            calibrated I magnitudes in different elliptical annuli 
            centred on NGC 1600. The solid line shows the
            source counts in the field region, away from NGC 1600 and 
            from the other galaxies in its group. The semi-major axis
            range (in pixels) of the different annuli is given in the 
            lower-right.
            Panels {\it b,c)}: The symbols are field-subtracted 
            counts in the same annuli as in panel {\it a}.
            The middle lines are Gaussian fits to the counts, using
            Eqs. (1) and (2). The lines immediately above and below
            each fit are Gaussian functions with a 1 $\sigma$
            variation in normalization relative to the fitted one.}
    \label{lfs1600i}
    \end{figure}

\section{Luminosity functions}

   Source counts as a function of B and I magnitudes were then derived 
   for several independent elliptical rings around NGC 1600. These are 
   shown in Figs. \ref{lfs1600b} and \ref{lfs1600i} for
   the samples selected from the B and I images, respectively. The units are 
   sources kpc$^{-2}$ and the plots are in logarithmic scale. Since a bin size
   of $\Delta mag = 0.5$ was used, the numbers were doubled to be normalized to unit magnitude. Image surface was
   converted into physical surface assuming that NGC 1600 is at 
   a distance of 63.0 Mpc (taken from {\it NASA Extragalactic
   Database, NED}\footnote[1]{http://nedwww.ipac.caltech.edu/}).This 
   corresponds to a distance modulus of $(m-M)_0 = 34.0$ and yields 
   a conversion of $1 kpc = 3.3 arcsec = 21 pix$. 
   The radial ranges given in the figure panels correspond to the inner 
   and outer semi-major axis 
   of each elliptical annulus used. Panel {\it a} 
   shows the total counts, including the background sources located
   away from NGC 1600 and the other galaxies in its group. The 
   remaining panels show the excess source counts around NGC 1600
   after the field counts were subtracted at each elliptical annulus.

   We assume that all excess counts are star clusters, most
   of them globular clusters, belonging to NGC 1600 and
   attempt to fit the globular cluster luminosity function (GCLF) at the 
   different radii. A Gaussian GCLF was assumed to have the form:

   $$\Phi(m) = \rho_0 e^{-{ {(m-\overline{m})^2} \over {2\sigma^2} } }, \eqno (1)$$

   \noindent where we assumed that $\overline{M_V} = -7.4$ and $\sigma=1.4$. This is
   an intermediate value between that quoted by Harris 1991 and that 
   used by Larsen et al (2001) based on HST data. 
   Typical GC colours of $(B-V)=0.85$ and $(V-I)=1.10$ were adopted 
   when fitting the functions to the B and I counts shown in the 
   figures. These are colours of old ($\tau > 10$ Grys) single stellar 
   populations with sub-solar ($Z \simeq 0.008$) metallicity, according to
   Bruzual \& Charlot (2003). By fixing the turn-over GCLF luminosity and
   its width, the fits had only the normalization, $\rho_0$, as 
   a free parameter. The most accurate Gaussian fits are represented 
   by the middle lines in Figs. \ref{lfs1600b}
   and \ref{lfs1600i}. They were obtained by varying the normalization 
   $\rho_0$ until the quantity

   $$\chi^2 = \Sigma_i^N { {(log \Phi - log \Phi_{obs})^2} \over {\delta log \Phi_{obs} }^2 }, \eqno (2)$$

   \noindent reached a minimum. In the expression above, $N$ was the number of points
   to be fitted in each region, $\Phi(m)$ was the Gaussian function given
   by eq. (1), $\Phi_{obs}$ is the observed source density. The 
   associated density fluctuation, $\delta log \Phi_{obs} \propto N_{obs}^{-1/2}$, 
   was found by propagating the Poisson fluctuation in the source number 
   counts, $N_{obs}$, in a given region and magnitude bin.

   The upper and lower lines shown in Figs. \ref{lfs1600b}
   and \ref{lfs1600i} represent $\pm 1 \sigma$ confidence levels 
   in $\rho_0$ for the best-fitting Gaussian functions. The variation
   in $\rho_0$ was measured 
   with 300 bootstrap realizations in which the source counts were varied 
   according to a Poisson distribution and the fit was repeated. In generating
   these realizations, a Gaussian distribution of shifts in the assumed distance 
   modulus and $\overline {M_V}$, with zero mean and standard deviation 
   $\sigma_{(m-M)}=0.1$ and $\sigma_{M_V} = 0.1$
   were also applied, to simulate uncertainties in these quantities.
   We emphasize that the adopted typical variation of 0.1 mag in
   distance modulus is larger than the expected uncertainties in the
   Hubble constant and flow models, and that this variation alters our
   answers by less than the amplitude of the Poisson fluctuations.

   These fits in conjunction with the assumed Gaussian model for 
   the GCLF allow us to estimate the size of the star cluster system of 
   NGC 1600.

   The $\rho_0$ values and the expected total number of GCs 
   are listed in Tables \ref{GCdensB} and \ref{GCdensI} for each annulus. 
   The latter is given by 

   $$N_{GC}[a_1,a_2] = \pi~{{b} \over {a} }~(a_2^2 - a_1^2)~\int_{-\infty}^{\infty}~\Phi(m)~dm = \pi~{{b} \over {a} }~(a_2^2 - a_1^2)~\rho_0~\sigma~\sqrt{2\pi}, \eqno (3)$$

   \noindent where $b/a$ is the axis ratio of the annulus, $a_1$ and $a_2$ are 
   the inner and outer semi-major axes, and $\Phi(m)$ is the GCLF in units
   of sources kpc$^{-2}$, as given in Eq. (1). The axis ratio is
   obtained from the isophote fitting of NGC 1600; it is $b/a=0.65$ with 
   little variation with position. Given the range $[a_1,a_2]$ and the
   fitted normalization density, we can derive the value of $N_{GC}[a_1,a_2]$ 
   with no additional free parameters. The uncertainties in $\rho_0$ shown in 
   Tables \ref{GCdensB} and \ref{GCdensI} are the 1 $\sigma$ spread in this quantity
   evaluated from the results of the
   300 bootstrap realizations described earlier. The uncertainties
   in $N_{GC}$ are propagated from those in $\rho_0$.

   \begin{table}
      \caption[]{Estimated star cluster surface densities at different 
                elliptical annuli from sample in B filter.}
         \label{GCdensB}
     $$ 
         \begin{array}{p{0.2\linewidth}p{0.10\linewidth}p{0.10\linewidth}p{0.15\linewidth}p{0.15\linewidth}p{0.15\linewidth}}
            \hline
            \noalign{\smallskip}
            Range (pix) & NPTS & NPTS$_w$ & $\rho_0 (kpc^{-2})$ & N$_{GC}$(range) & N$_{GC}$ (cum.) \\
            \noalign{\smallskip}
            \hline
            \noalign{\smallskip}
            0-100 & * & * & 1.59 $\pm$ 0.40 & 253.5 $\pm$ 63.8 & 253.5 $\pm$ 63.8 \\
            100-200 & 36 & 85.4 & 0.82 $\pm$ 0.21 & 392.2 $\pm$ 100.4 & 645.7 $\pm$ 164.2 \\
            200-350 & 121 & 267.2 & 0.60 $\pm$ 0.09 & 789.2 $\pm$ 118.4 & 1434.9 $\pm$ 282.6 \\
            350-500 & 175 & 378.5 & 0.38 $\pm$ 0.05 & 772.5 $\pm$ 101.6 & 2207.4 $\pm$ 384.2 \\
            500-800 & 413 & 761.2 & 0.08 $\pm$ 0.02 & 497.4 $\pm$ 124.4 & 2704.8 $\pm$ 508.6 \\
            \noalign{\smallskip}
            \hline
         \end{array}
     $$
\begin{list}{}{}
\item[] Column 1: range in pixels between inner and outer semi-major
axes of annulus used for source counts (1 kpc = 21 pixels); 
Column 2: total number of sources found in the annulus; Column 3: total 
weighted number of sources in annulus; Column 4: normalization density of 
Gaussian fit to GCLF at given annulus, as explained in text (in units of 
$kpc^{-2}$); total expected number of clusters in annulus, 
obtained as explained in the text. 
\end{list}
\end{table}

\begin{table}
      \caption[]{Estimated star cluster surface densities at different 
            elliptical annuli from sample in I filter.}
         \label{GCdensI}
     $$ 
         \begin{array}{p{0.2\linewidth}p{0.10\linewidth}p{0.10\linewidth}p{0.15\linewidth}p{0.15\linewidth}p{0.15\linewidth}}
            \hline
            \noalign{\smallskip}
            Range (pix) & NPTS & NPTS$_w$ & $\rho_0 (kpc^{-2})$ & N$_{GC}$(range) & N$_{GC}$ (cum.)\\
            \noalign{\smallskip}
            \hline
            \noalign{\smallskip}
            0-150 & * & * & 2.20 $\pm$ 0.44 & 789.2 $\pm$ 157.8 & 789.2 $\pm$ 157.8 \\
            150-250 & 79 & 156.0 & 1.47 $\pm$ 0.30 & 937.5 $\pm$ 191.3 & 1726.7 $\pm$ 349.1 \\
            250-350 & 155 & 323.4 & 0.37 $\pm$ 0.06 & 354.0 $\pm$ 57.4 & 2080.7 $\pm$ 406.5 \\
            350-500 & 191 & 353.2 & 0.27 $\pm$ 0.04 & 548.9 $\pm$ 81.3 & 2629.6 $\pm$ 487.8 \\
            500-650 & 163 & 357.6 & 0.13 $\pm$ 0.02 & 357.5 $\pm$ 55.0 & 2987.1 $\pm$ 542.8 \\
            \noalign{\smallskip}
            \hline
         \end{array}
     $$
\begin{list}{}{}
\item[] Column 1: range in pixels between inner and outer semi-major 
axes of annulus used for source counts (1 kpc = 21 pixels); Column 2: 
total number of sources found in the annulus; Column 3: total 
weighted number of sources in annulus; Column 4: normalization density of 
Gaussian fit to GCLF at given annulus, as explained in text (in units of 
$kpc^{-2}$); total expected number of clusters in annulus, obtained as 
explained in the text. 
\end{list}
\end{table}

   \begin{figure}
   \centering
   \includegraphics[width=\textwidth]{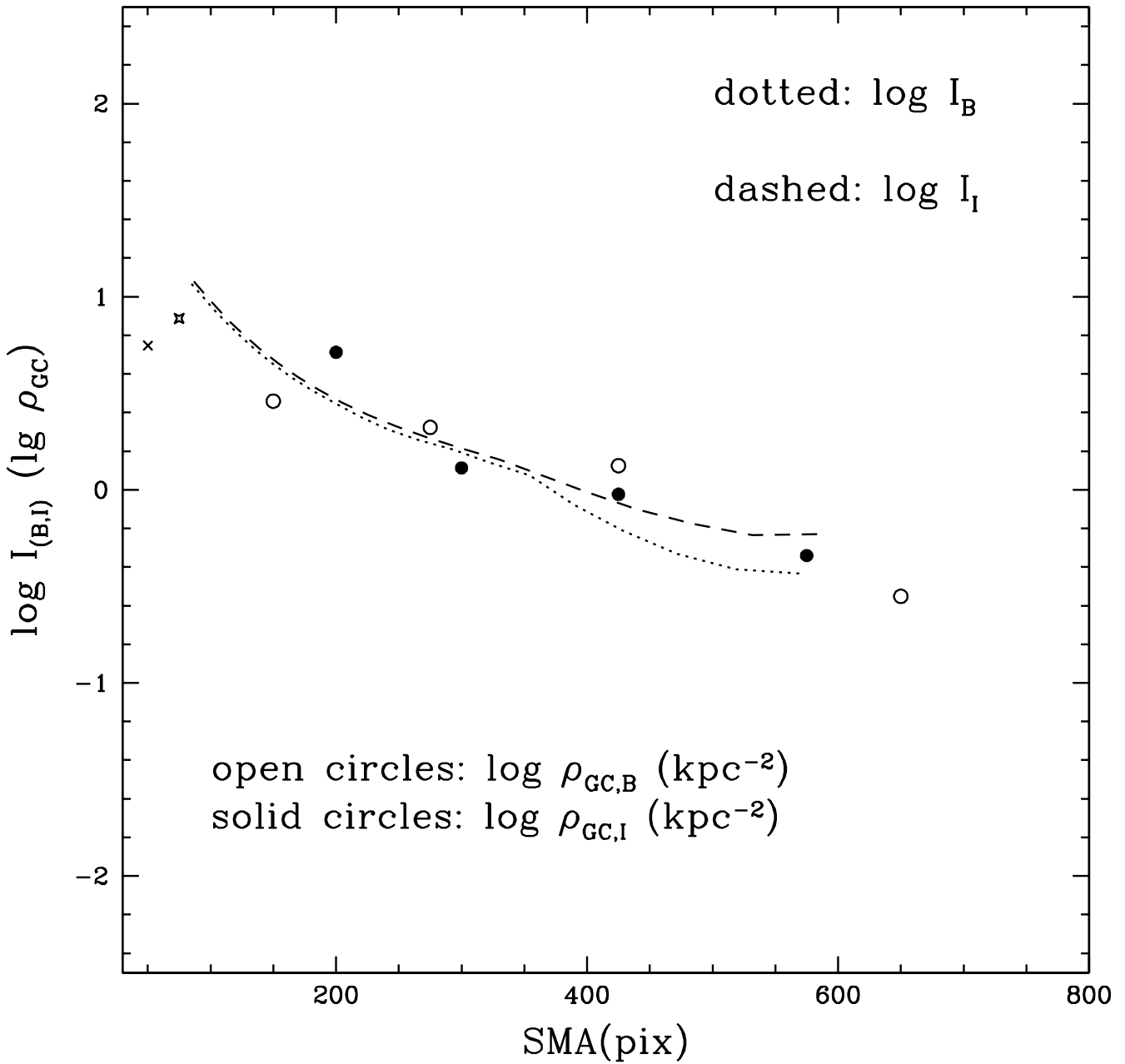}
   \caption{The points correspond to the star cluster density profile 
            measured 
            using source counts in the I (solid circles) and B bands 
            (open circles). The star and the cross in the inner region 
            are linear extrapolations in the figure of the measured 
            densities in I and B, respectively. The curves show the 
            surface brightness 
            profiles in I (dashed line) and B (dotted line), in arbitrary 
            units.}
    \label{profs}
    \end{figure}

   The points marked with an asterisk in Tables \ref{GCdensB} and 
   \ref{GCdensI} are estimates based on inward extrapolations
   of the source densities from the other bins. They correspond to
   regions in which the high intensity of the galaxy caused saturation 
   (or a non-linear response regime on the CCD) and prevented reliable detection 
   and measurement of magnitudes for a sizable sample. 
   The study of these inner regions is however
   necessary to allow an estimate of the total number of star clusters in
   NGC 1600. Also shown in the tables is the cumulative
   number of clusters. The system size derived from these counts
   is almost independent of the band used. The expected total number
   of clusters in NGC 1600 is $N_{GC} = (2.85 \pm 0.50)~10^3$. 
   Assuming $M_V=-23.1$ to be the absolute magnitude of this galaxy, 
   we derive $S_N = 1.6 \pm 0.3$
   for its specific frequency of clusters. This is relatively low
   for a luminous elliptical such as NGC 1600. Dirsch et al 2003 measured  
   $S_N = 5.1 \pm 1.2$ for the cD in Fornax cluster, NGC 1399. Rhode \& Zepf 
   (2004) found that $S_N = 3.5 \pm 0.5$ and $S_N = 3.6 \pm 0.6$ for NGC 4406 and
   NGC 4472, respectively, in Virgo. These galaxies are of similar luminosity but located 
   in higher density environments than NGC 1600. It is therefore possible that the 
   difference in $S_N$ is an environmental effect. However, 
   Forbes et al (1997) measured $S_N = 4.3 \pm 1.1$ for the luminous cD group 
   galaxy NGC 5846. Another 
   possibility is that the star cluster system of NGC 1600 extends 
   further beyond the edge of the SOAR image field-of-view, even though the field includes all  
   diffuse light from the galaxy. We cannot presently exclude this 
   possibility and this $S_N$ estimate should be regarded as a lower limit.

   \begin{figure}
   \centering
   \includegraphics[width=\textwidth]{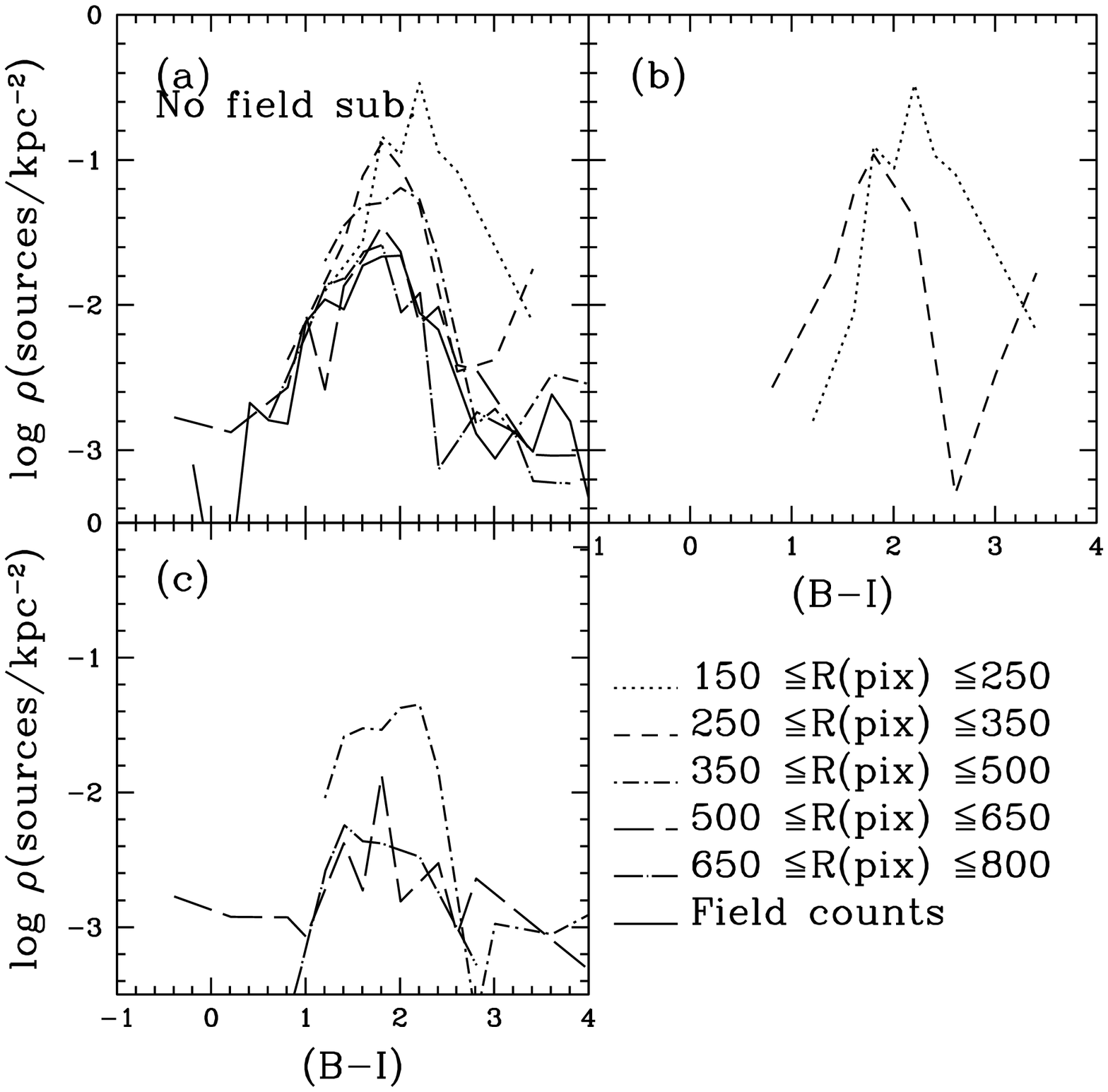}
   \caption{Panel {\it a)}: Total source counts as a function of calibrated 
            (B-I) colours in different elliptical annuli 
            centred on NGC 1600. The solid line shows the source counts
            in the field region, away from any of the galaxies in NGC1600 
            group. Panels {\it b,c)}: Field subtracted counts in
            the same annuli as in panel {\it a)}, as indicated.}
    \label{colhist}
    \end{figure}

   Figure \ref{profs} shows the estimated density of stellar 
   clusters belonging to NGC 1600 as a function of galactocentric 
   distance, expressed in terms of semi-major axis of the elliptical
   annulli. The densities are in units of star clusters kpc$^{-2}$ and were 
   obtained by integration of Eq. (1) over all magnitudes at each 
   distance bin. The points shown as stars or crosses are the result of 
   inward extrapolation of the measured densities, as discussed above.
   Figure \ref{profs} also shows, for comparison, the surface brightness profiles of 
   NGC 1600 for both filters. These profiles were derived  
   from the isophotal fitting of data for NGC 1600 described earlier. We note  
   that the cluster system follows the light profile of the field 
   stars out to the projected distances probed ($R \leq 800$ pix $\simeq 
   40$ kpc). This is in contrast to that observed for several other
   luminous ellipticals (Harris 1986), although, as afore mentioned, 
   we cannot exclude the existence of an envelope of star clusters 
   outside the field studied.

\section{Colour distributions}

    Colour distributions are a useful discriminator between the different 
    star cluster sub-populations, and may reflect different star and
    cluster formation events within a galaxy. Figure \ref{colhist} shows
    the $(B-I)$ colour distribution for the point sources that have 
    colour information. The units are again in sources kpc$^{-2}$ and
    the bin size used was $\Delta (B-I) = 0.2$. 
    Panel {\it a} shows total colour counts
    of sources with $B < 27$ for the same annuli as in Fig. 
    \ref{lfs1600i}. Panels {\it b} and {\it c} show the field subtracted
    source counts.

    A spatial colour gradient is visible in the panels, in the
    sense that redder star clusters tend to be found closer to the centre
    of the host galaxy. The average star cluster colour varies from
    $\overline{(B-I)}=2.2$ in the innermost region shown in Fig. 
    \ref{colhist} to $\overline{(B-I)}=1.9$ in the $350 \leq R \leq 500$ 
    pix ($17 \leq R \leq 25$ kpc) annulus. This may reflect true changes 
    in metallicity
    or age, although redenning caused by dust may also contribute. Similar
    trends were observed by Forbes et al (1997) in the central group 
    elliptical NGC 5846, and are also common in luminous
    cluster ellipticals (Lee \& Geisler 1993, Harris et al 1998, 
    Larsen et al 2001).
    There is a hint of bimodality in the distributions associated with 
    NGC 1600 both in the inner regions and beyond $R \simeq 350$ 
    pix ($R > 17$ kpc). 
    The large peak in the inner regions of NGC 1600, with $(B-I) 
    \simeq 2.2$ (see panel {\it b}), is typical of an old ($\tau > 10 
    Gyrs$) and fairly metal-rich ($Z \simeq Z_{\odot} = 0.019$) single 
    stellar population 
    according to the models by Bruzual \& Charlot (2003) and assuming
    extinction to be negligible. Comparable $(B-I)$ colours were found  
    by Grillmair et al (1999) for NGC 1399 and NGC 1404. The smaller
    peak at $(B-I) \simeq 1.8$ is also
    present in the region $250 \leq R \leq 350$ pix ($12 \leq R
    \leq 16$ kpc), again shown in panel {\it b} of Fig. \ref{colhist}. 
    This colour requires lower metallicity $Z \leq 0.004$ for an old 
    population or significantly younger ages, $\tau \simeq 2.5$ Gyrs, at 
    solar abundance.
    
    A clearly larger spread in colours, with two likely peaks, 
    is observed in the intermediate annulus $350 \leq R \leq 500$ pix ($17 
    \leq R \leq 25$ kpc), shown in panel {\it c} of Fig.
    \ref{colhist}. The redder peak has approximately 
    the same colour, $(B-I) \simeq 2.0$, as an old star 
    cluster with $Z \simeq 0.008$. The bluer peak, however, has 
    $(B-I) \simeq 
    1.6$. According to Fig. \ref{profs}, this region coincides with
    the position of a bump in the light profile of the galaxy, which is 
    more pronounced in the blue. Probably, this is a ring or shell. Its blue colour is 
    also consistent with the
    smooth profile in the near infra-red bands shown by Rembold 
    et al (2002). In fact, Forbes \& Thomson (1992) identified
    an outer shell in NGC 1600 in an R band image.
    It is possible that the larger spread in colours about this distance
    may be associated with this feature and the bluer cluster peak 
    may represent a younger population of star clusters 
    formed at the onset of the perturbation in the host galaxy.

   \begin{figure}
   \centering
   \includegraphics[width=\textwidth]{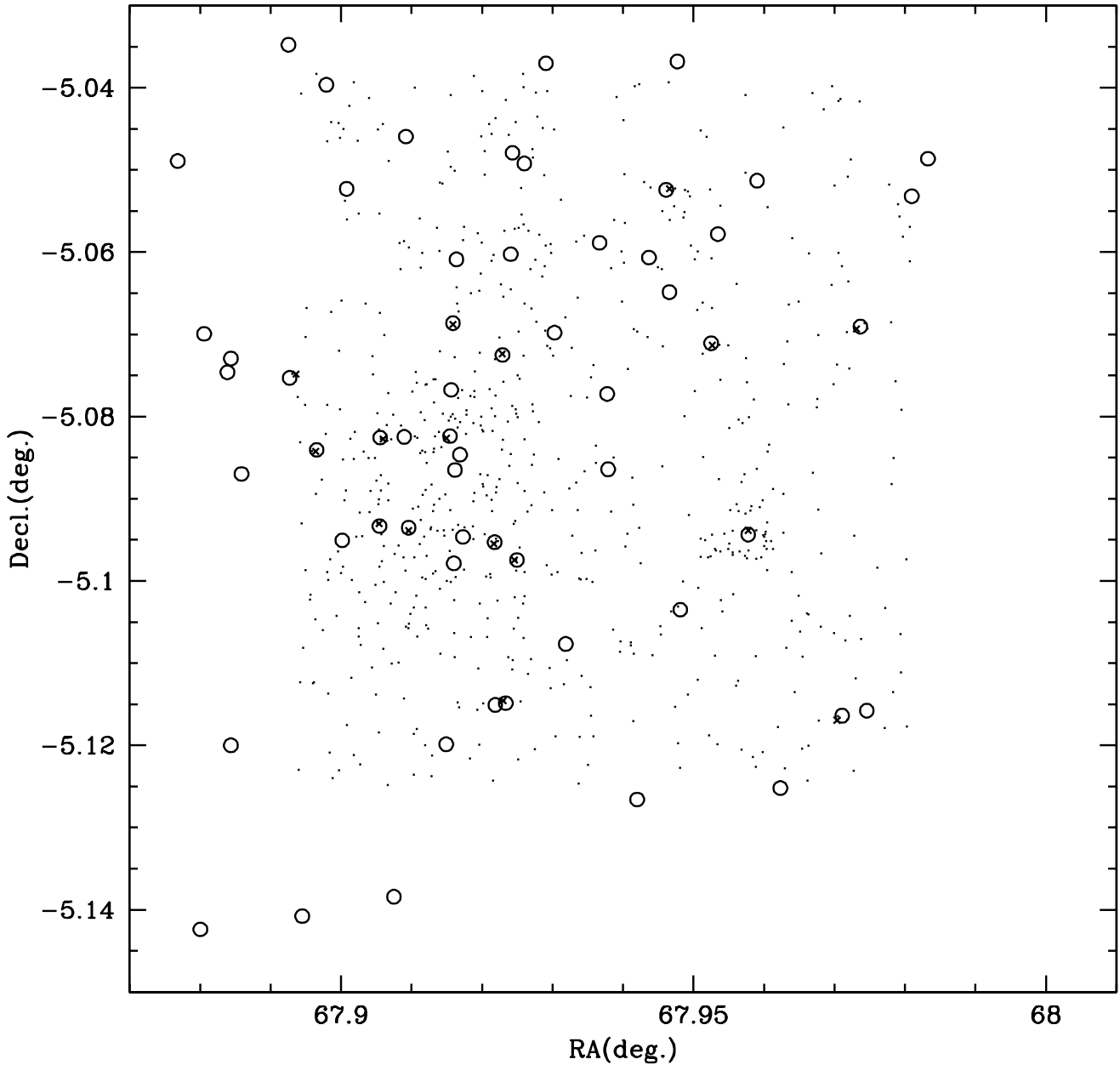}
   \caption{On-sky distribution of optical and X-ray sources in NGC1600's 
            group. The small dots are star cluster candidates selected in
            this work. Only sources with magnitudes and colours in the 
            ranges where an excess is seen relative to the field are
            are plotted. The large circles are the discrete X-ray sources
            from Sivakoff et al (2004). The crosses indicate matched
            optical/X-ray sources.}
    \label{xraymat}
    \end{figure}

\section{X-ray counterpars}

The sample of star cluster candidates was cross-referenced with the list 
of 71 discrete X-ray sources from Sivakoff et al (2004). Positional matches 
were achieved on the basis of the equatorial coordinates. The astrometry 
of the SOI sources assumed a linear relation between CCD and on-sky
positions, for the CCD plate scale of 0.154 arcsec/pixel. This assumption was
tested with the known coordinates of the 3 galaxies imaged, NGC 1600, NGC 1603,
and NGC 1601. The differences between their equatorial coordinates, as 
listed in the NASA Extragalactic Database (NED), was compared with the 
expected differences based on the adopted linear transformation from 
their CCD (x,y) positions. Discrepancies of $< 1.5$ arcsec were found in 
all cases, both in declination and right ascension. 

In attempting to identify the cluster candidates with the X-ray sources 
we permitted residual offsets in our coordinates relative to those listed
by Sivakoff et al (2004). The approach was to start with the 
null-hypothesis that several optical/X-ray coincidences should be found, and 
then search for the offsets in right ascension and declination, 
within 3 arcsec from our nominal positions, that maximized the number of 
matches between the two samples. Only optical sources, whose magnitudes
and colours were consistent with those of star clusters in NGC 1600 
(or in its group), were used in this positional matching process. 

Of the 71 discrete X-ray sources identified by Sivakoff et al (2004), a total of 
49 were located inside the SOI field limits. Sixteen optical/X-ray coincidences 
were found, and the offsets in equatorial coordinates that resulted in this number 
of matches were 1.75arcsec in right ascension and 0.25arcsec in declination.
Figure \ref{xraymat} shows the on-sky distribution of the optical and
X-ray sources. The two clumps of optical sources correspond to NGC 1600 and
NGC 1603. Eleven of the coincidences are clearly associated with NGC
1600, being within 40 kpc (132 arcsec) of its centre. A total of 28 X-ray
sources are in this same region, yielding a coincidence rate of 39\%.
This may be an underestimate, given the optical sample incompleteness, 
although there is a known trend towards optical/X-ray coincidences being 
related to bright sources in the optical 
(Chies-Santos et al 2006, Kundu et al 2007, Sivakoff et al 2007). 
For example, Kundu et al (2007) studied 5 early-type galaxies using 
HST data, and found that the vast majority of clusters harbouring low-mass 
X-ray binaries (LMXBs) have $M_I < -9$. Similarly, most GCs with a LMXB 
in the sample of 11 luminous elliptical and S0 galaxies in Virgo studied by 
Sivakoff et al (2007) have $M_z < -9$.
Due to the detection limits of the SOAR photometry, we note that 
all 16 matched sources in this work have $M_I < -10$ (or $M_V < -8.8$)
if they are at the distance of the NGC 1600 group.
 
Kundu et al (2007) and Sivakoff et al (2007) also found that GCs with LMXBs had
a strong tendency to 
belong to the red peak of the colour distribution. In the 
case of NGC 1399, Kundu et al (2007) found that all clusters with LXMBs had 
$(B-I) > 1.7$. In our NGC 1600 sample, we found 4 optical/X-ray matches 
with bluer $(B-I)$ values. The
remaining 7 objects have redder colours. Since the sample was small and we are
unable to confirm that individual objects are star clusters in 
NGC 1600, it is currently impossible to assess if this apparent discrepancy 
is significant.

Table \ref{matches} lists the 16 sources found to coincide with
X-ray discrete sources; both optical and X-ray properties, the latter 
taken from Sivakoff et al (2004), are listed. The top 11 entries are those
of sources located around NGC 1600. Inspection of Table 
\ref{matches} does not reveal any strong systematics between X-ray
luminosity and optical colour or magnitude, although the range in the
latter quantity is small. Only one of the sources listed in the table 
was previously related to an optical counterpart by Sivakoff
et al (2004): this was source 40, which is close to the centre of the
companion galaxy NGC 1603. All sources matched but one, source 3, 
are consistent with being point sources, according to these authors.
Source 3 is close to the centre of NGC 1600.

   \begin{table}
      \caption[]{Sample of Optical/X-ray coincidences.}
         \label{matches}
     $$ 
         \begin{array}{p{0.1\linewidth}p{0.3\linewidth}p{0.3\linewidth}p{0.1\linewidth}p{0.1\linewidth}p{0.1\linewidth}}
            \hline
            \noalign{\smallskip}
            Source & $\alpha$ & $\delta$ & $L_X$ & $B$ & $(B-I)$ \\
            \noalign{\smallskip}
            \hline
            \noalign{\smallskip}
            3 & 04:31:39.71 & -05:04:56.6 & 30.1 & 24.85 & 1.82 \\
            7 & 04:31:38.30 & -05:05:36.7 & 30.8 & 24.56 & 2.10 \\
            8 & 04:31:41.23 & -05:05:43.0 & 19.7 & 24.21 & 1.56 \\
            9 & 04:31:37.33 & -05:04:57.2 & 14.3 & 24.40 & 2.11 \\
            11 & 04:31:37.31 & -05:05:36.0 & 5.9 & 24.59 & 1.87 \\
            12 & 04:31:41.99 & -05:05:50.8 & 8.4 & 25.08 & 1.80 \\
            13 & 04:31:41.50 & -05:04:21.0 & 6.3 & 24.34 & 1.56 \\
            14 & 04:31:39.81 & -05:04:07.1 & 15.0 & 24.58 & 1.94 \\
            16 & 04:31:35.17 & -05:05:02.6 & 39.0 & 24.24 & 1.57 \\
            21 & 04:31:34.25 & -05:04:31.1 & 22.8 & 24.53 & 2.29 \\
            25 & 04:31:41.61 & -05:06:53.5 & 52.0 & 24.29 & 1.44 \\
            38 & 04:31:48.60 & -05:04:15.9 & 5.5 & 24.49 & 1.85 \\
            40 & 04:31:49.86 & -05:05:39.7 & 13.0 & 21.94 & 1.93 \\
            41 & 04:31:47.07 & -05:03:08.7 & 11.9 & 24.58 & 1.35 \\
            55 & 04:31:53.68 & -05:04:08.6 & 108.3 & 24.69 & 2.14 \\
            59 & 04:31:53.06 & -05:06:59.0 & 19.2 & 23.78 & 2.56 \\
            \noalign{\smallskip}
            \hline
         \end{array}
     $$
\begin{list}{}{}
\item[] Column 1: source identification number given by Table 2 of 
Sivakoff et al (2004); Columns 2 and 3: equatorial coordinates (J2000)
of X-ray sources, given by Sivakoff et al (2004); column 4: X-ray 
luminosity in units of $10^{38}$ ergs/s; columns 5 and 6: calibrated $B$ 
magnitude and $(B-I)$ colour from this work.
\end{list}
\end{table}

\section{Conclusions}

Deep B and I photometry of the galaxy group of NGC 1600 has been obtained
using the SOAR telescope. Several thousand point sources were found
in the field for both filters. There was a significant excess of
point sources around the luminous elliptical NGC 1600, which we
associated with its star cluster system. Magnitudes and colours were measured
and sample incompleteness was also quantified. The star cluster 
luminosity functions at different elliptical annuli around NGC 1600 were 
all consistent with a Gaussian of typical average luminosity and 
dispersion. Assuming this functional form, the total surface 
density of star clusters, corrected for sampling incompleteness, was obtained
as a function of galactocentric distance. The density distribution was
similar to the underlying light distribution of the galaxy.
A total of $N_{GC} = 2850 \pm 500$ star clusters were estimated to exist 
within the central $40$ kpc of NGC 1600, yielding a specific frequency 
of $S_N=1.6 \pm 0.3$ in this region. Since the cluster system may extend 
beyond the diffuse galaxy light and the limits of the field studied, this
estimate should be considered as a lower limit.

The colour distribution had a hint of bimodality, especially at the
centre of NGC 1600 and at an intermediate region $\simeq 20$ kpc
away. The reddest and most significant peak at the centre is consistent with 
a $\tau > 10$ Gyr old metal-rich ($Z \simeq Z_{\odot}$) star 
cluster population. The bluer peak, which also
dominates the cluster counts at $R \simeq 10-15$ kpc, is either an 
old population with 1/5 of the solar abundance or a $2.5$ Gyr old 
population of solar metallicity. 

The colour distribution became particularly wide and bimodal
at $\simeq 20$ kpc from the centre of NGC 1600. This region coincided
with a bump in the surface brightness profile of the host galaxy,
which is clear in the $B$ filter image and completely
absent in the near infra-red. This is probably a ring or shell,
as previously suggested by Forbes \& Thomson (1992). This  
perturbation may have been accompanied by star and cluster formation. The
bluer colours found in the cluster sample in this region may
be related to such a star formation event, which would indicate a
relatively young star cluster population.

Finally, the optical cluster sample was cross-correlated with the discrete X-ray
sources from Sivakoff et al (2004). The coincidence rate within
40 kpc of the galaxy centre, where an excess of point
sources in the optical is clearly seen, is $\simeq 40\%$. The optical/X-ray
coincidences vary widely in X-ray luminosity and optical
colours, but due to the magnitude limit of the SOAR image they all
correspond to optically luminous star clusters ($M_I < -10$). Since previous 
studies, based on Chandra and HST data, indicated that clusters with LMXBs are 
preferentially luminous, it is possible that the sample of optical/X-ray
matches that we have found is a reasonably complete census of GCs harbouring X-ray
sources in NGC 1600. 
Except for one source, all matched sources are consistent
with being unresolved in the Chandra images.

\begin{acknowledgements}
      This work was supported by Conselho Nacional de Desenvolvimento
      Cient\'\i fico e Tecnol\'ogico (CNPq) in Brazil through a research
      grant to BXS. The author thanks
      the staff at SOAR for taking the data.
\end{acknowledgements}


\begin{thebibliography}{}


  \bibitem[1992]{ash92} Ashman, K., \& Zepf, S., 1992, ApJ, 384, 50.

  \bibitem[1984]{bar84} Barbon, R., Benacchio, L., Capaccioli, M., \&
  Rampazzo, R., 1984, A\&A, 137, 166.

  \bibitem[2002]{bro02} Brodie, J., \& Larsen, S., 2002, AJ, 124, 1410.

  \bibitem[2006]{bro06} Brodie, J., \& Strader, J., 2006, ARA\&A, 44, 193.

  \bibitem[2003]{bru03} Bruzual, G., \& Charlot, S., 2003, MNRAS, 344, 1000.

  \bibitem[1982]{bur82} Burstein, D., \& Heiles, C., 1982, AJ, 87, 1165.

  \bibitem[2006]{chies06} Chies-Santos, A., Pastoriza, M., Santiago, B., \& 
		    Forbes, D., 2006, A\&A, 455, 453

  \bibitem[2007]{chies07} Chies-Santos, A., Pastoriza, M., \& Santiago, B., 2007, A\&A, 467, 1003

  \bibitem[1998]{cote98} Cot\^e, P., Marzke, R., \& West, M., 1998, 501, 554

  \bibitem[2003]{dir03} Dirsch, B., Richtler, T., Geisler, D., Forte, J., Bassino, L., \& Gieren, W., 2003, AJ, 125, 1908

  \bibitem[1996]{elson96} Elson, R., \& Santiago, B., 1996, MNRAS, 280, 971.

  \bibitem[1992]{forb92} Forbes, D., \& Thomson, R., 1992, MNRAS, 254, 723

  \bibitem[1997]{forb97} Forbes, D., Brodie, J., \& Huchra, J., 1997, AJ, 113, 887.

  \bibitem[1999]{gri99} Grillmair, C.J., Forbes, D., Brodie, J.P., \&  Elson, R.A.W., 1999, 117, 167

  \bibitem[1986]{har86} Harris, W., 1986, AJ, 91, 822.

  \bibitem[1991]{har91} Harris, W., 1991, ARA\&A, 29, 543.

  \bibitem[1998]{har98} Harris, W., Harris, G., \& McLaughlin, D.,  1998, AJ, 115, 1801

  \bibitem[2006]{har06} Harris, W., Whitmore, B., Karakla, D., Oko\'n,
  W., Baum, W., et al, 2006, ApJ, 636, 90.

  \bibitem[1997]{kiss97} Kissler-Patig, M., Kohle, S., Hilker, M.,
		Richtler, T., Infante, L. \& Quintana, H., 1997, A\&A, 
                319, 470.

  \bibitem[1999]{kun99} Kundu, A, Whitmore, B., Sparks, W., Macchetto,
		D., Zepf, S., Ashman, K., 1999, ApJ, 513, 733.

  \bibitem[2001]{kun01} Kundu, A, \& Whitmore, B., 2001, AJ, 121, 2950.

  \bibitem[2007]{kun07} Kundu, A., Maccarone, T., \& Zepf, S., 2007, ApJ, 662, 525.

  \bibitem[1992]{land92} Landolt, A., 1992, AJ, 104, 340.

  \bibitem[2000]{lar00} Larsen, S, \& Brodie, J., 2000, AJ, 120, 2938.

  \bibitem[2001]{lar01} Larsen, S, Brodie, J., Huchra, J., Forbes, D., \& Grilmair, C., 2001, AJ, 121, 2974.

  \bibitem[1993]{lee93} Lee, M.G., \& Geisler, D., 1993, AJ, 106, 493

  \bibitem[1995]{mah95} Mahabal, A., Kembhavi, A., Singh, K., \& Green,
  R., 1995, JApAS, 16, 206.

  \bibitem[2006]{peng06a} Peng, E., Jord\'an, A., Cot\'e, P., et al,
  2006, ApJ, 639, 95.

  \bibitem[2006]{peng06b} Peng, E., Cot\'e, P., Jord\'an, A., et al,
  2006, ApJ, 639, 838.

  \bibitem[2004]{rhode04} Rhode, K. \& Zepf, S., 2004, AJ, 127, 302

  \bibitem[2002]{remb02} Rembold, S., Pastoriza, M., Ducati, J., Rubio,
  M., Roth, M., 2002, A\&A, 391, 531

  \bibitem[1998]{sch98} Schlegel, D., Finkbeiner, D., \& Davis, M.,
  1998, ApJ, 500, 525.

  \bibitem[2004]{siv04} Sivakoff, G., Sarazin, C., \& Carlin, J., 2004,
  ApJ, 617, 262. 

  \bibitem[2007]{siv07} Sivakoff, G., Jord\'an, A., Sarazin, C., Blakeslee, J., C\^ot\'e, P., Ferrarese, L., Juett, A., Mei, S., \& Peng, E., 2007, ApJ, 660, 1246.

  \bibitem[2006]{stra06} Strader, J., Brodie, J., Spitler, L., \& Beasley,
  M., 2006, AJ, 132, 2333.

\end{thebibliography}
\end{document}